# Generalized coordinate transformations for Monte Carlo (DOSXYZnrc and VMC++) verifications of DICOM compatible radiotherapy treatment plans

**Richard M. Schmitz[1,2], Oliver Telfer[1,2], Reid W. Townson[1,2] and Sergei Zavgorodni[2,1]**

[1] Department of Physics and Astronomy, University of Victoria, Victoria, British Columbia, Canada
[2] Department of Medical Physics, BC Cancer Agency, Vancouver Island Centre, Victoria, British Columbia, Canada

Corresponding author: Sergei Zavgorodni (szavgorodni@bccancer.bc.ca)

**Abstract.** The International Electrotechnical Commission (IEC) has previously defined standard rotation operators for positive gantry, collimator and couch rotations for the radiotherapy DICOM coordinate system that is commonly used by treatment planning systems. Coordinate transformations to the coordinate systems of commonly used Monte Carlo (MC) codes (BEAMnrc/DOSXYZnrc and VMC++) have been derived and published in the literature. However, these coordinate transformations disregard patient orientation during the computed tomography (CT) scan, and assume the most commonly used "head first, supine" orientation. While less common, other patient orientations are used in clinics – Monte Carlo verification of such treatments can be problematic due to the lack of appropriate coordinate transformations.

In this work, a solution has been obtained by correcting the CT-derived phantom orientation and deriving generalized coordinate transformations for field angles in the DOSXYZnrc and VMC++ codes. The rotation operator that includes any possible patient treatment orientation was determined using the DICOM *Image Orientation* tag (0020,0037). The derived transformations of the patient image and beam direction angles were verified by comparison of MC dose distributions with the Eclipse[TM] treatment planning system, calculated for each of the eight possible patient orientations.

## 1. Introduction

Monte Carlo (MC) verification of radiation therapy treatment plans has become a widely accepted method of benchmarking new commercial dose calculation algorithms. Additionally, MC calculations are beginning to play a role in clinical quality assurance. Commonly employed Monte Carlo "dose calculation engines" include DOSXYZnrc (Walters, Kawrakow *et al.*, 2009) and VMC++ (Kawrakow, 2001). The coordinate system used in the DICOM standard is defined by the patient: the positive x-axis points toward the patient's left-hand side, the positive y-axis points toward the patient's posterior and the positive z-axis points from inferior to superior direction (figure 1(a) and (b)). The International Electrotechnical Commission (IEC) standard defines coordinate systems for radiotherapy treatment machines and rotation operators for transformation to and from the DICOM patient coordinate system (International Electrotechnical Commission 2007). DOSXYZnrc (isource=2) uses a spherical polar-coordinate system based on the DICOM coordinate system, while VMC++ (with phase-space source) uses the standard Euler angles between the particle transport code BEAMnrc and DICOM coordinate systems.

Coordinate transformations from the DICOM coordinate system to that of the MC transport code are required to describe the radiation beams incident on a phantom. At least three separate derivations of these transformations have been published (Thebaut and Zavgorodni, 2006, Thebaut and Zavgorodni, 2007, Bush and Zavgorodni, 2010, Zhan, Jiang *et al.*, 2012). However, these methods assume that the patient is oriented in the head-first supine position. In general, there are eight possible positions for the patient to lie in on a treatment bed: head-first and feet-first variants of supine, prone, decubitus-left and decubitus-right. Therefore, an extra rotation of the coordinate systems is required to correctly account for patient orientation. Additionally, the *Image Position* tag of the DICOM image must be considered and corrected for in some cases by an extra image rotation and translation.

In this paper, a method to account for patient position at imaging is presented. The method consists of two stages: the image is rotated and translated to bring it to "head-first-supine" orientation, and then a rotation matrix is applied to the beam direction angles to correctly orient beams for MC codes. The coordinate transformations have been derived for two commonly used radiotherapy dose calculation codes: DOSXYZnrc and VMC++.

## 2. Materials and Methods

### 2.1. Patient orientation matrix

The DICOM standard (DICOM, 2011) provides a description of the direction cosines for the image rows and columns through the *Image Orientation* tag (0020,0037), which contains two vectors:

$$\boldsymbol{v}_r = \begin{pmatrix} v_{r1} \\ v_{r2} \\ 0 \end{pmatrix} \tag{1}$$

$$\boldsymbol{v}_c = \begin{pmatrix} v_{c1} \\ v_{c2} \\ 0 \end{pmatrix} \tag{2}$$

The vector $\boldsymbol{v}_r$ specifies the direction of the first row of the image with respect to the patient coordinate system. Similarly, the vector $\boldsymbol{v}_c$ specifies the direction of the first column of the image (DICOM, 2011). Restrictions are imposed on these direction vectors by the physical limitations of the patient's orientation (e.g. the patient may not be standing up or be lying perpendicular to the couch). This forces the z-component of $\boldsymbol{v}_r$ and $\boldsymbol{v}_c$ to be zero. In other words, each image is a cross-section in the patient's XY-plane. All possible values of $\boldsymbol{v}_r$ and $\boldsymbol{v}_c$ are given in Table 1. The normal vector of the image, $\boldsymbol{v}_n$ , is then the cross product of these two, and *x* and *y* components equal to zero:

$$\boldsymbol{v}_n = \boldsymbol{v}_r \times \boldsymbol{v}_c = \begin{pmatrix} 0 \\ 0 \\ v_{r1}v_{c2} - v_{r2}v_{c1} \end{pmatrix} \tag{3}$$

We can now define a rotation matrix *P* to be used to describe patient orientation in the DICOM coordinate system. Since $\boldsymbol{v}_r$ and $\boldsymbol{v}_c$ describe the rows and columns of the image (i.e. the usual *x* and *y* unit vectors) in the basis of the patient's rotated orientation, *P* is determined as the matrix with $\boldsymbol{v}_r$, $\boldsymbol{v}_c$ and $\boldsymbol{v}_n$ as its columns:

$$P \overset{\text{def}}{=} \begin{bmatrix} P_{11} & P_{12} & 0 \\ P_{21} & P_{22} & 0 \\ 0 & 0 & P_{33} \end{bmatrix} = \begin{bmatrix} v_{r1} & v_{c1} & 0 \\ v_{r2} & v_{c2} & 0 \\ 0 & 0 & v_{r1}v_{c2} - v_{r2}v_{c1} \end{bmatrix} \tag{4}$$

It is worthwhile to note that $P$ is an orthonormal matrix with a determinant of unity, and that all $P_{ij}$ are either -1, 0 or 1.

**Table 1**[*]: A list of possible patient positions and their image orientations as found in the DICOM tag (0020,0037).

| Patient Position | Image Orientation Vectors |
|---|---|
| **Head-first supine (HFS)** | $\mathbf{v}_r = <1, 0, 0> \mathbf{v}_c = <0, 1, 0>$ |
| **Head-first decubitus-left (HFDL)** | $\mathbf{v}_r = <0, -1, 0> \mathbf{v}_c = <1, 0, 0>$ |
| **Head-first decubitus-right (HFDR)** | $\mathbf{v}_r = <0, 1, 0> \mathbf{v}_c = <-1, 0, 0>$ |
| **Head-first prone (HFP)** | $\mathbf{v}_r = <-1, 0, 0> \mathbf{v}_c = <0, -1, 0>$ |
| **Feet-first supine (FFS)\*** | $\mathbf{v}_r = <-1, 0, 0> \mathbf{v}_c = <0, 1, 0>$ |
| **Feet-first decubitus-left (FFDL)\*** | $\mathbf{v}_r = <0, 1, 0> \mathbf{v}_c = <1, 0, 0>$ |
| **Feet-first decubitus-right (FFDR)\*** | $\mathbf{v}_r = <0, -1, 0> \mathbf{v}_c = <-1, 0, 0>$ |
| **Feet-first prone (FFP)\*** | $\mathbf{v}_r = <1, 0, 0> \mathbf{v}_c = <0, -1, 0>$ |

[*] Vector values in Table 1 assume that the patient images have not been altered. Images from patients scanned in these positions may be automatically altered by the treatment planning software to appear in the head-first variants of these positions. This would alter image orientation vectors.

### 2.2. Rotation and Translation of the Phantom

There are three DICOM tags that must be used to determine the correct positioning of the patient (DICOM, 2011). These are the *Image Position (Patient)* tag, the *Image Orientation* tag and the *Patient Position* tag (summarized in Table 2).

**Table 2:** Summary of DICOM tags necessary to report patient positioning during imaging and treatment.

| DICOM Tag Name | DICOM Tag ID | Description |
|---|---|---|
| Image Position (Patient) | (0020,0032) | *(x, y)* location of the top-left corner of this image in the DICOM coordinate system |
| Image Orientation | (0020,0037) | Two vectors, $v_r$ and $v_c$, containing the direction cosines for the first row and first column of this image with respect to a head-first supine DICOM coordinate system. |
| Patient Position | (0018,5100) | A 3- or 4- letter code describing the orientation in which the patient was scanned. For example, may contain "HFS" for "head-first supine". |

When creating a virtual patient phantom from DICOM images using the ctcreate code (Walters, Kawrakow *et al.*, 2009), the patient orientation is by default assumed to be "head-first, supine". To create a virtual phantom for patients treated in any other orientation, the *Image Orientation* (0020,0037) tag needs to be taken into account and extra image transformations will be required to correctly orient the phantom relative to the beam isocentre. If these transformations are omitted, the resulting phantom will be orientated and located incorrectly. For example, a patient scanned in head-first supine would have an *Image Position* (0020,0032) tag in quadrant III of the XY-plane and the image would be generated from

Generealized transformations for Monte Carlo verifications of radiotherapy treatment plans this location going in the positive $x$ and $y$ directions (see figure 2(a)). In contrast, a head-first prone patient has an *Image Position* tag in quadrant I and the image should be generated going in the negative $x$ and $y$ directions according to the direction cosines given by $\boldsymbol{v}_r$ and $\boldsymbol{v}_c$ (see Table 1). Instead, if the *Image Position* tag is disregarded, the image is still generated going in the positive $x$ and $y$ directions, resulting in a phantom that is rotated by 180° about its centre and translated away from the origin (figure 2 (b), lower-right). Since the isocentre location and direction of the beams are still unaltered, this results in the simulated beams aligning improperly with respect to the patient geometry.

To correct this, a simple algorithm was implemented that rotates the phantom around the DICOM z-axis using the rotation matrix $P$ and then repositions it accordingly to be in the standard head-first supine position with respect to the DICOM coordinate system. For HFP patients, the phantom was translated left and up by the image width and height, respectively (figure 2 (b)). For HFDR patients, the phantom was translated left by the image height. For HFDL patients, the phantom was translated up by the image width.

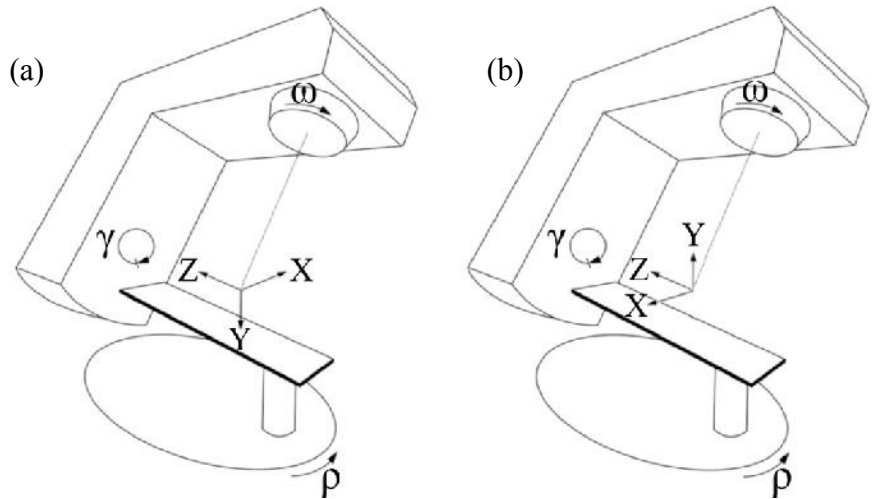

**Figure 1:** Orientation of the patient coordinate system for head-first supine and head-first prone positions are given in (a) and (b), respectively.

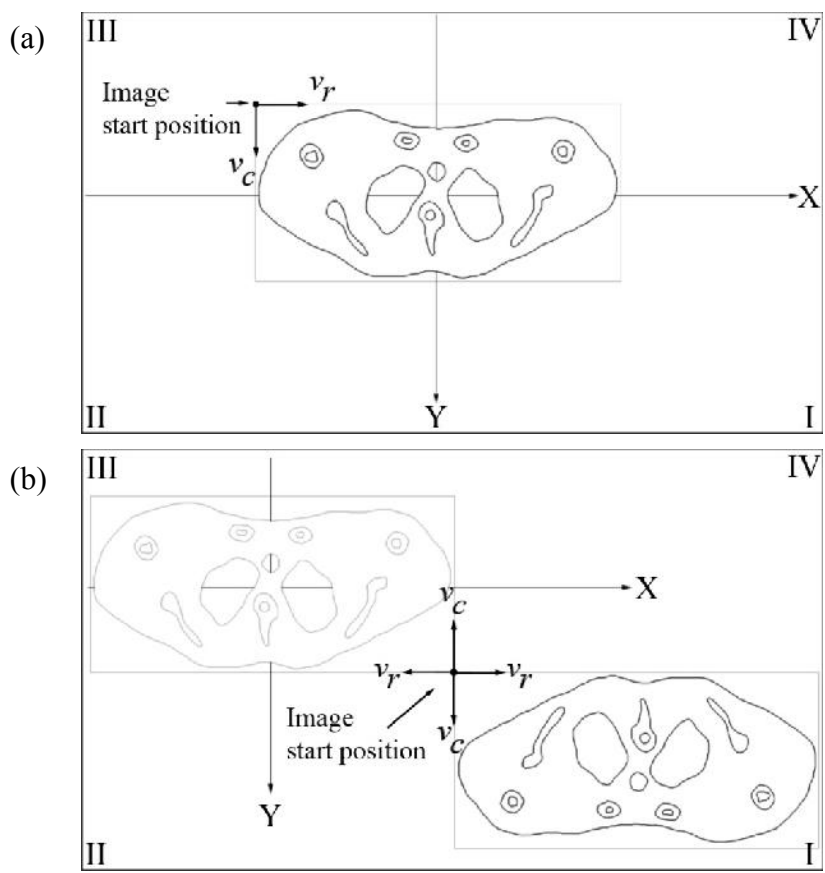

**Figure 2:** Figure (a) shows a cross section in the XY-plane of a phantom created by ctcreate from a patient scanned in the head-first supine position. Its image position coordinate is in quadrant III. Figure (b, lower-right) shows the phantom of a patient scanned in the head-first prone position. This phantom is misplaced and rotated. Figure (b, upper-left) illustrates the rotation and translation of the image required to obtain a correctly oriented phantom.

### 2.3. Additional Rotation for Feet-First Patients

The CT scanning software used in our clinic automatically reoriented the patient images into a head-first position, such that all *Image Orientation* vectors were of the head-first type. This presented a problem when computing the transformation matrix $T'$, described below, as it is dependent on having unaltered vectors $v_r$ and $v_c$. To correct for this problem, all images having a *Patient Position* tag (0018,5100) beginning with FF (for "feet-first") and whose normal vector $v_n$ was not equal to -1 (which indicates feet-first) were subject to a rotation by 180° about the *y*-axis before rotation by the matrix $P$. Rotation by 180° about the *y*-axis orients the DICOM coordinate system in a feet-first position.

### 2.4. General Coordinate Transformation for Radiation Fields in DOSXYZnrc

DOSXYZnrc requires the angle of beam incidence to be described using the polar-coordinate angles $\theta$ and $\varphi$, where $\theta$ is the polar angle in the DICOM coordinate system having a domain from 0 to $\pi$ radians and $\varphi$ is the azimuthal angle with a domain from 0 to $2\pi$ radians (Walters, Kawrakow *et al.*, 2009). The rotation of the beam about its central axis is described by $\varphi_{coll}$. Thebaut and Zavgorodni derived equations

Generealized transformations for Monte Carlo verifications of radiotherapy treatment plans for these angles (Thebaut and Zavgorodni, 2006, Thebaut and Zavgorodni, 2007) using the transformation matrix $T$:

$$T = \begin{bmatrix} \sin\rho & \cos\gamma\cos\rho & \sin\gamma\cos\rho \\ 0 & \sin\gamma & -\cos\gamma \\ -\cos\rho & \cos\gamma\sin\rho & \sin\gamma\sin\rho \end{bmatrix} \tag{5}$$

Here, $\gamma$, $\rho$ and are the linac gantry and couch angles, respectively.

In order to remove assumption of the patient head-first supine position, the new coordinate transformation matrix $T'$ that is the result of matrix multiplication between the former $T$ and $P$ should be used:

$$T' = PT = \begin{bmatrix} P_{11}\sin\rho & P_{11}\cos\gamma\cos\rho + P_{12}\sin\gamma & P_{11}\sin\gamma\cos\rho - P_{12}\cos\gamma \\ P_{21}\sin\rho & P_{21}\cos\gamma\cos\rho + P_{22}\sin\gamma & P_{21}\sin\gamma\cos\rho - P_{22}\cos\gamma \\ -P_{33}\cos\rho & P_{33}\cos\gamma\sin\rho & P_{33}\sin\gamma\sin\rho \end{bmatrix} \tag{6}$$

Using equation (6) and following the same technique as described by Thebaut and Zavgorodni (Thebaut and Zavgorodni, 2006) we can now derive a general set of equations for $\theta$, $\varphi$ and $\varphi_{coll}$ that accounts for the patient orientation.

In the linac coordinate system the beam direction vector is a constant vector (Thebaut and Zavgorodni, 2006), namely:

$$\boldsymbol{beam}_{linac} = \begin{pmatrix} 0 \\ 0 \\ -1 \end{pmatrix} \tag{7}$$

Therefore in the DICOM coordinate system we have,

$$\boldsymbol{beam}_{DICOM} = T'\boldsymbol{beam}_{linac} = \begin{pmatrix} P_{12}\cos\gamma - P_{11}\sin\gamma\cos\rho \\ P_{22}\cos\gamma - P_{21}\sin\gamma\cos\rho \\ -P_{33}\sin\gamma\sin\rho \end{pmatrix} \overset{\text{def}}{=} \begin{pmatrix} X_{beam} \\ Y_{beam} \\ Z_{beam} \end{pmatrix} \tag{8}$$

Our new equations are as follows:

$$\theta \overset{\text{def}}{=} \cos^{-1}(-Z_{beam}) = \cos^{-1}(P_{33}\sin\gamma\sin\rho) \tag{9}$$

$$\varphi \overset{\text{def}}{=} \tan^{-1}\left(\frac{Y_{beam}}{X_{beam}}\right) = \tan^{-1}\left(\frac{P_{22}\cos\gamma - P_{21}\sin\gamma\cos\rho}{P_{12}\cos\gamma - P_{11}\sin\gamma\cos\rho}\right) \tag{10}$$

Care must be taken when calculating $\varphi$ to use a two-argument arctangent function. This increases the range of the function to a full 360°, allowing the correct angle to be computed. In order to derive the quantities $\varphi_{coll}$ and *direction*, we define the vectors $\boldsymbol{Coll}_{DICOM}$ and $\boldsymbol{L}_{DICOM}$ and use them as in Thebaut and Zavgorodni:

$$\boldsymbol{Coll}_{DICOM} \stackrel{\text{def}}{=} T' \begin{pmatrix} \sin\omega \\ -\cos\omega \\ 0 \end{pmatrix}_{linac} \tag{11}$$

$$\boldsymbol{L}_{DICOM} \stackrel{\text{def}}{=} \begin{pmatrix} -\sin\varphi \\ \cos\varphi \\ 0 \end{pmatrix}_{DICOM} \tag{12}$$

Then the equations for $|\varphi_{coll}|$ and *direction* may be obtained as:

$$|\varphi_{coll}| \stackrel{\text{def}}{=} \cos^{-1}(\boldsymbol{Coll}_{DICOM} \cdot \boldsymbol{L}_{DICOM}) \tag{13}$$

$$direction \stackrel{\text{def}}{=} (\boldsymbol{Coll}_{DICOM} \times \boldsymbol{L}_{DICOM}) \cdot \boldsymbol{beam}_{DICOM} \tag{14}$$

*2.5. General Coordinate Transformations for Radiation Fields in VMC++*

VMC++ modeling requires the incident beam angle to be described in standard Euler angles (α, β, γ) representing the orientation of the linac coordinate system relative to the DICOM coordinate system. Rotation of an arbitrary vector by an angle *q* around the X, Y and Z axes of the patient DICOM coordinate system can be described by rotation operators $R_X$, $R_Y$, $R_Z$ as:

$$R_X(\theta) = \begin{bmatrix} 1 & 0 & 0 \\ 0 & \cos\theta & -\sin\theta \\ 0 & \sin\theta & \cos\theta \end{bmatrix} \tag{15}$$

$$R_Y(\theta) = \begin{bmatrix} \cos\theta & 0 & \sin\theta \\ 0 & 1 & 0 \\ -\sin\theta & 0 & \cos\theta \end{bmatrix} \tag{16}$$

$$R_Z(\theta) = \begin{bmatrix} \cos\theta & -\sin\theta & 0 \\ \sin\theta & \cos\theta & 0 \\ 0 & 0 & 1 \end{bmatrix} \tag{17}$$

The rotation of the linac can described using rotation matrices for the non-commutative order of collimator, gantry, and couch as described by Bush and Zavgorodni (Bush and Zavgorodni, 2010):

$$R_{accel} = R_{-Y}(\theta_T)R_Z(\theta_G)R_{-Y}(\theta_C) \tag{18}$$

where $\theta_T$ is the angle of the couch, $\theta_G$ is the angle of the gantry and $\theta_C$ is the angle of the collimator. Here $R_Z$ is a positive rotation around the z-axis, and $R_{-Y}$ is a negative rotation around the y-axis. It should be noted that the normal vector of the phase space plane is initially aligned along the positive z-axis of the patient coordinate system. To align the phase space system with the zeroed position of the collimator, gantry and couch angles an additional rotation of -90° about the DICOM x-axis is needed. Therefore the amended relation is:

$$R_{accel} = R_{-Y}(\theta_T)R_Z(\theta_G)R_{-Y}(\theta_C)R_X\left(-\frac{\pi}{2}\right) \tag{19}$$

However, this only holds for the patient in the head-first supine position. For all other cases, as in the previous section, the rotation $P$ must also be applied to the DICOM rotation operator. The final operator that describes linac rotation in the DICOM coordinate system then becomes:

$$R_{accel} = PR_{-Y}(\theta_T)R_Z(\theta_G)R_{-Y}(\theta_C)R_X\left(-\frac{\pi}{2}\right) \tag{20}$$

These rotations can also be described in the VMC++ coordinate system using rotation matrices with the standard Euler angles:

$$R_{VMC++} = R_X(\alpha)R_Y(\beta)R_Z(\gamma) \tag{21}$$

Thus, we have two 3x3 matrices that are term-wise equal to each other:

$$R_{accel} = R_{VMC++} \tag{22}$$

or

$$PR_{-Y}(\theta_T)R_Z(\theta_G)R_{-Y}(\theta_C)R_X\left(-\frac{\pi}{2}\right) = R_X(\alpha)R_Y(\beta)R_Z(\gamma) \tag{23}$$

The solution for this equation is over-determined, and was obtained with mathematical computing software, giving:

$$\alpha = \tan^{-1}\left(\frac{P_{22}\cos\theta_G - P_{21}\sin\theta_T\sin\theta_G}{-P_{33}\sin\theta_T\sin\theta_G}\right) \tag{24}$$

$$\beta = -\tan^{-1}\left(\frac{P_{11}\cos\theta_T\sin\theta_G - P_{12}\cos\theta_G}{\sqrt{1 - (P_{11}\cos\theta_T\sin\theta_G - P_{12}\cos\theta_G)^2}}\right) \tag{25}$$

$$\gamma = \tan^{-1}\left(\frac{P_{11}\cos\theta_C\sin\theta_T + \sin\theta_C[P_{11}\cos\theta_T\cos\theta_G + P_{12}\sin\theta_G]}{\cos\theta_C[P_{11}\cos\theta_T\cos\theta_G + P_{12}\sin\theta_G] - P_{11}\sin\theta_T\sin\theta_C}\right) \tag{26}$$

### 2.6. Verification of the derived transformations

In order to verify the validity of the new transformation matrix $T'$, as well as the new equations for Euler angles used in VMC++, the derived equations were implemented into the Vancouver Island Monte Carlo (VIMC) system (Zavgorodni, Bush *et al.*, 2007, Bush, Townson *et al.*, 2008). A RANDO® phantom (The Phantom Laboratory) was scanned with all possible patient orientations and treatment plans were constructed in the Eclipse[TM] TPS (Varian Medical Systems, Palo Alta, Ca. USA). Beam directions (combinations of gantry, couch and collimator angles) were used that included all quadrants of gantry, couch and collimator rotations. Dose calculations were performed in Eclipse[TM] and compared to those done by the VIMC system. Treatment plans were tested for each of the eight possible patient orientations with MC calculations performed with approximately 1% uncertainty near the isocentre.

## 3. Results and Discussion

Table 3 summarizes gantry, couch and collimator angles for each plan and field tested. Figures 3 through 10 illustrate the agreement of the MC calculations with the TPS-calculated dose. Dose differences are within estimated statistical uncertainty – this confirms the validity of the derived transformations for all possible patient orientations.

Clinically, most patients are scanned and treated in the HFS position, and therefore these transformations will only be required for a relatively small data sub-set. However, with the development of new treatment techniques, the relative fraction of such treatments may increase. Currently, the most common non-HFS treatments at our centre are treatments in HFP position. These tend to include spine irradiations, prone breast techniques, and some pelvic treatments, as well as other sites. Other patient treatment orientations are relatively rare, but they can occur, and the generalized approach and transformations derived in this paper will be useful to perform MC verifications of such treatments.

**Table 1:** Summary of gantry, couch and collimator angles for each of the 8 test plans used.

| Patient Orientation(s) | Field Number | $\gamma$ (°) | $\rho$ (°) | $\omega$ (°) |
|:---:|:---:|:---:|:---:|:---:|
| **HFS** | 1 | 330.0 | 290.0 | 220.0 |
| | 2 | 40.0 | 50.0 | 165.0 |
| | 3 | 200.0 | 345.0 | 315.0 |
| **HFDL, HFDR, HFP, FFS, FFDL, FFDR, FFP** | 1 | 30.0 | 290.0 | 40.0 |
| | 2 | 320.0 | 50.0 | 345.0 |
| | 3 | 160.0 | 345.0 | 135.0 |

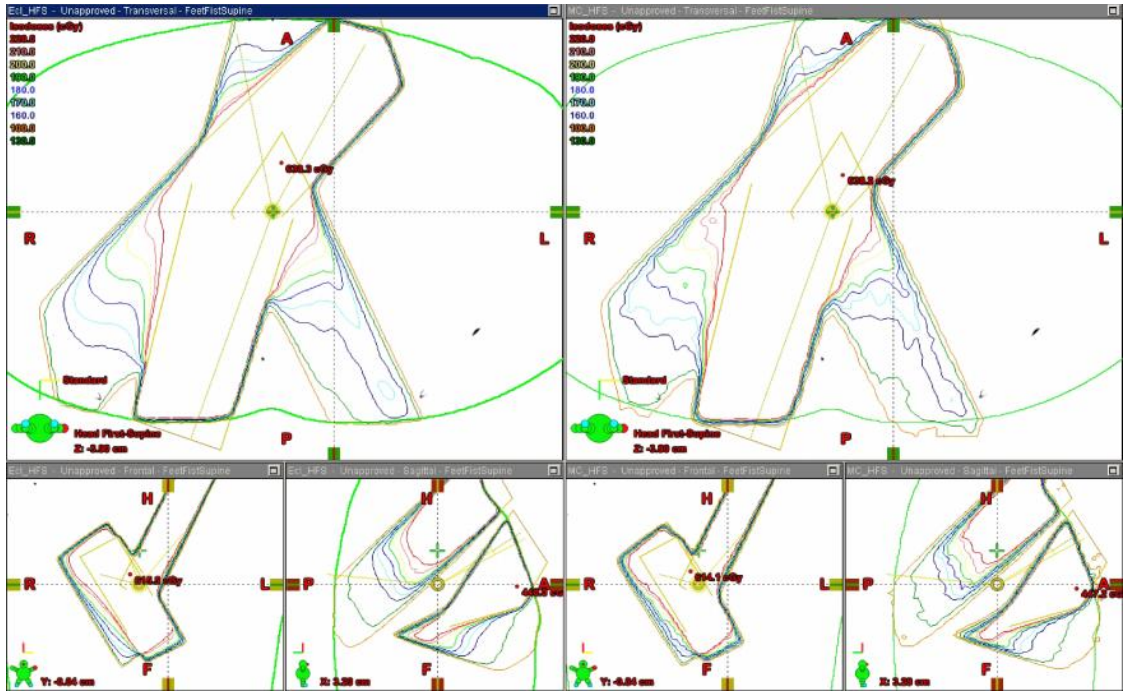

**Figure 3:** TPS (left) and Monte Carlo (right) calculated dose distribution for a patient scanned in the head-first supine position.

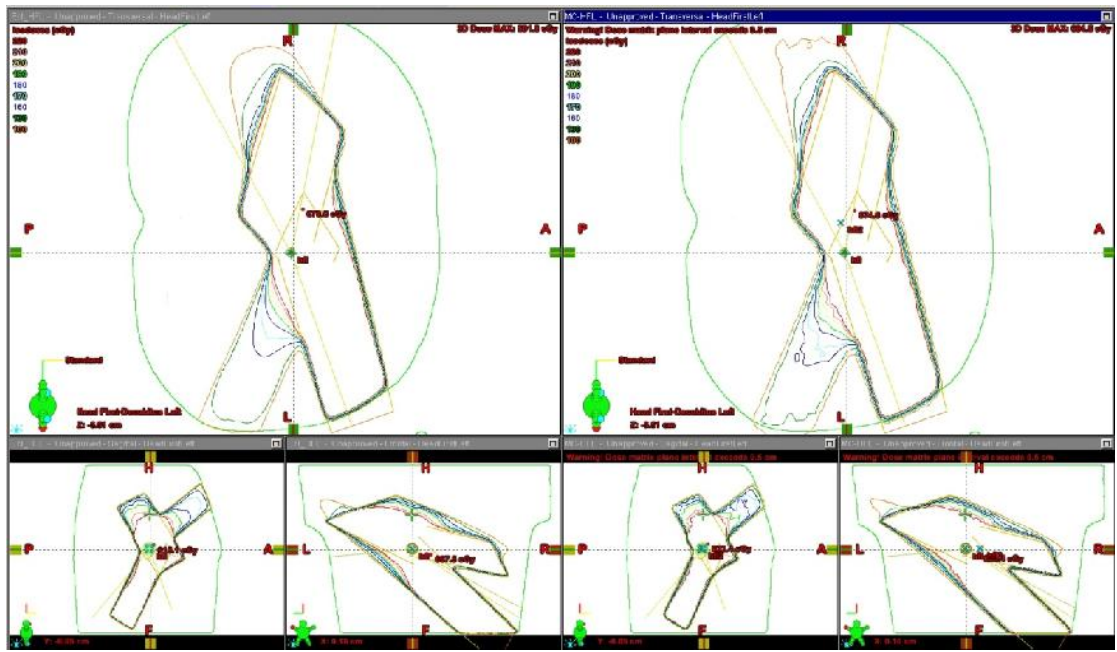

**Figure 4:** TPS (left) and Monte Carlo (right) calculated dose distribution for a patient scanned in the head-first decubitus-left position.

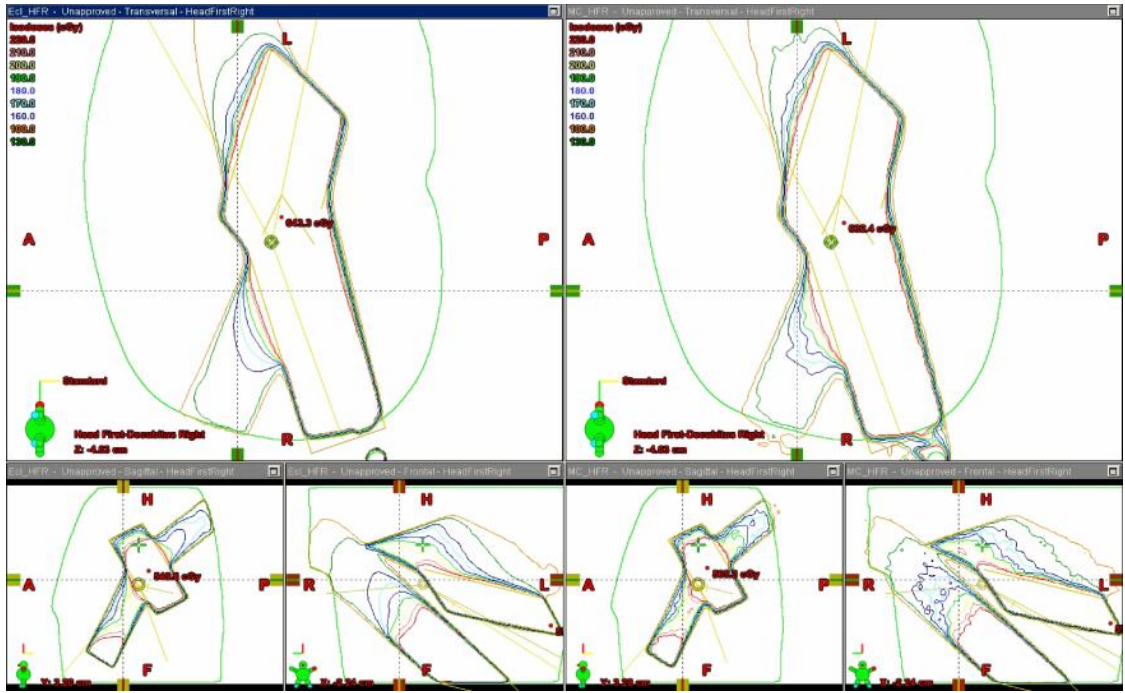

**Figure 5:** TPS (left) and Monte Carlo (right) calculated dose distribution for a patient scanned in the head-first decubitus-right position.

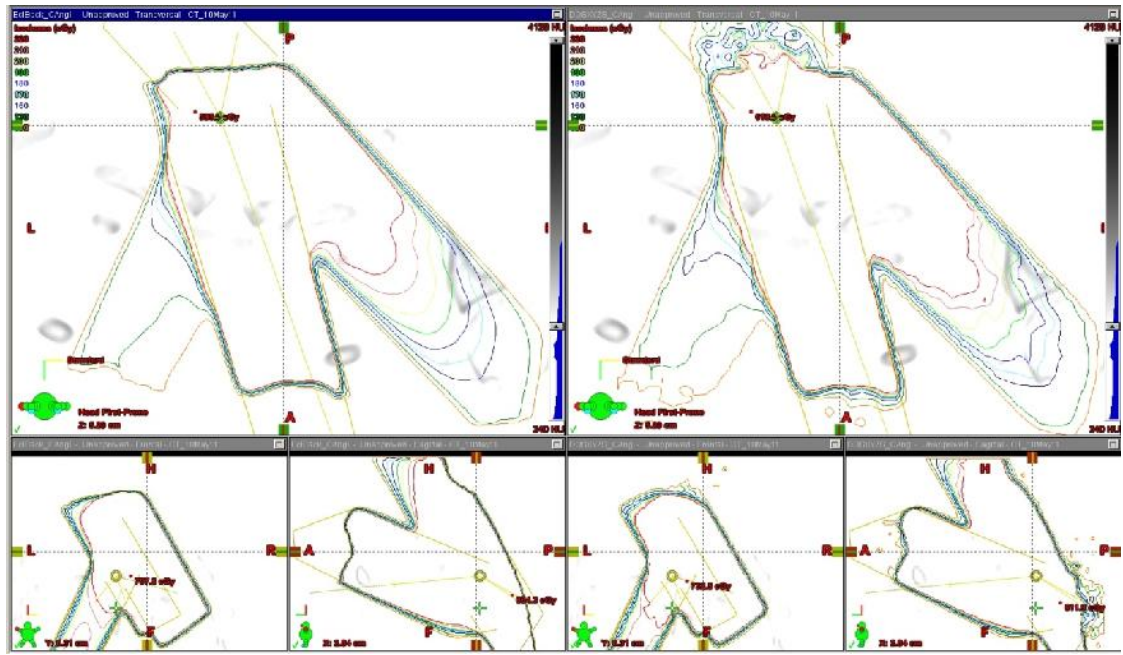

**Figure 6:** TPS (left) and Monte Carlo (right) calculated dose distribution for a patient scanned in the head-first prone position.

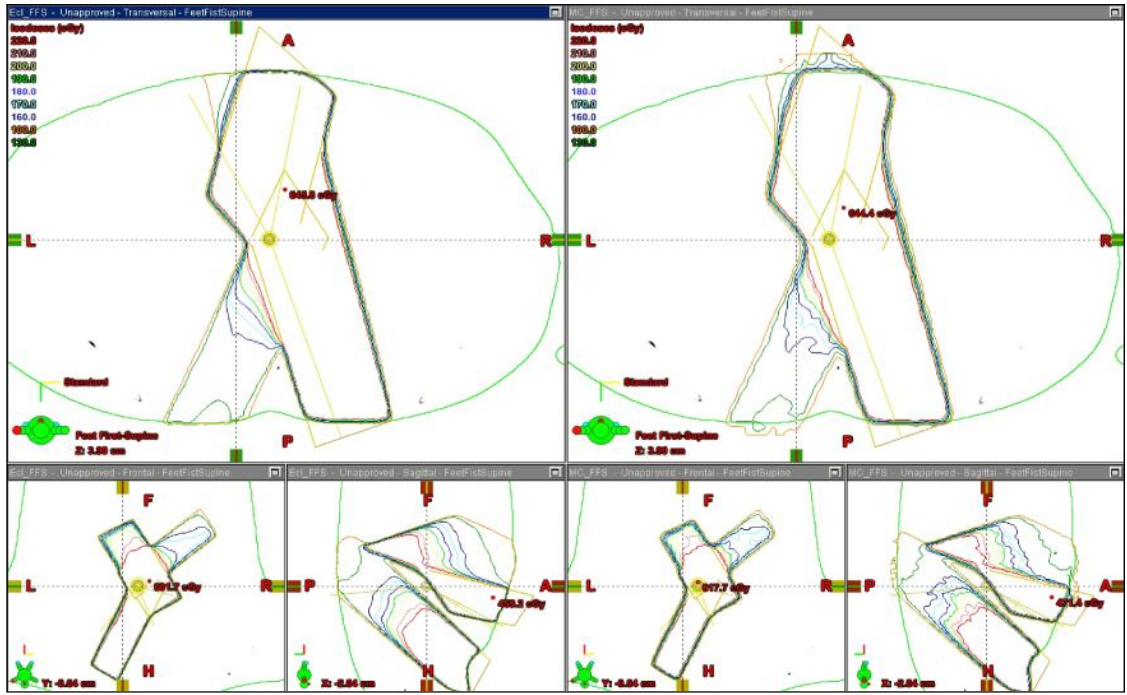

**Figure 7:** TPS (left) and Monte Carlo (right) calculated dose distribution for a patient scanned in the feet-first supine position.

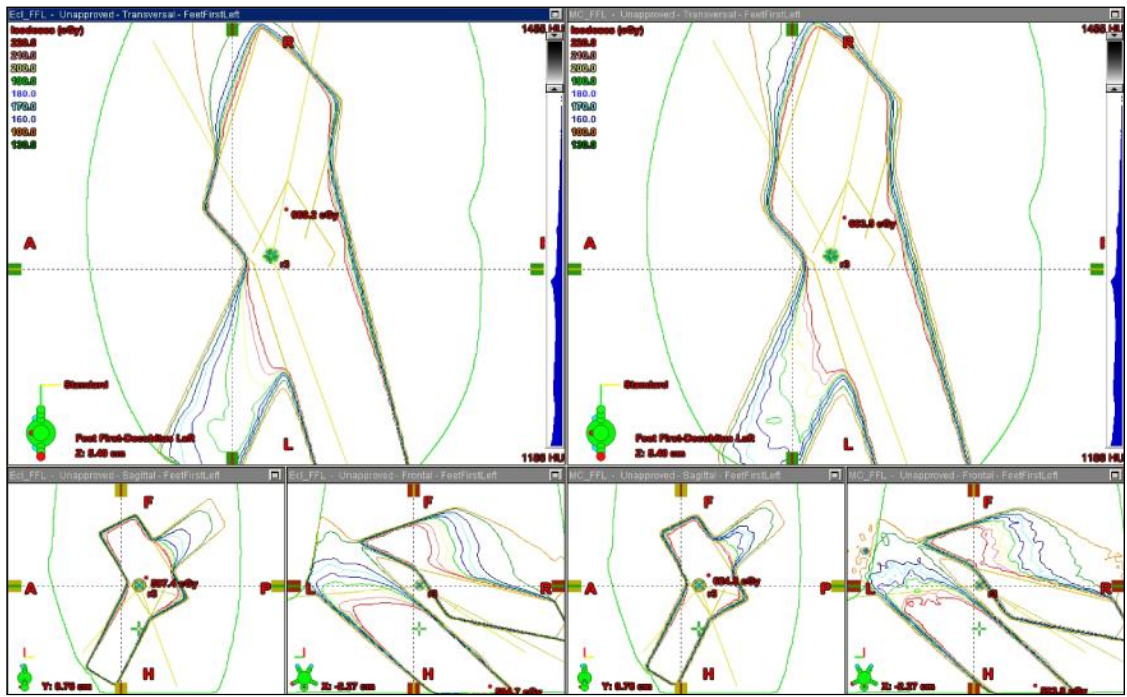

**Figure 8:** TPS (left) and Monte Carlo (right) calculated dose distribution for a patient scanned in the feet-first decubitus-left position.

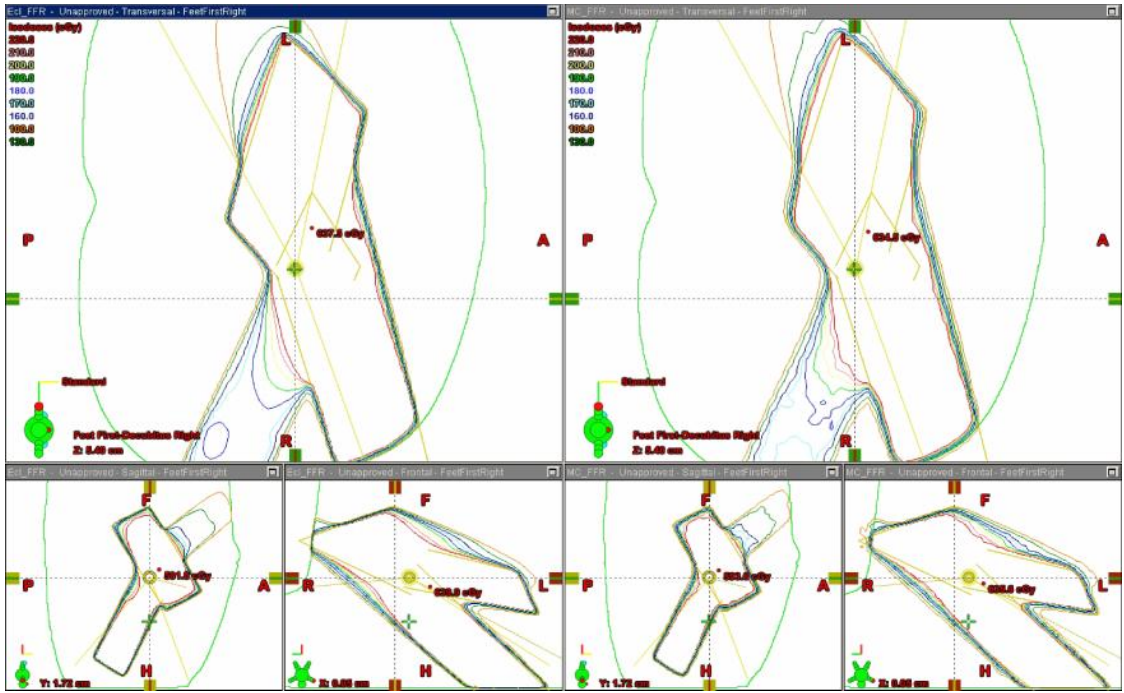

**Figure 9:** TPS (left) and Monte Carlo (right) calculated dose distribution for a patient scanned in the feet-first decubitus-right position.

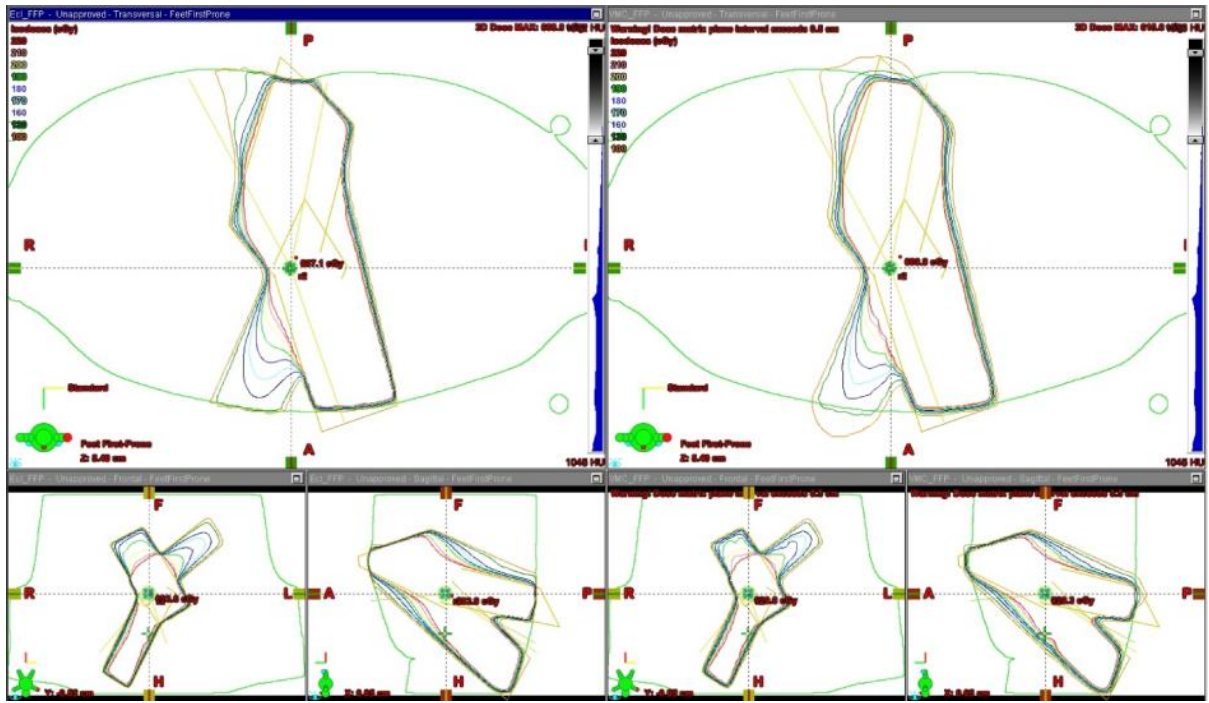

**Figure 10:** TPS (left) and Monte Carlo (right) calculated dose distribution for a patient scanned in the feet-first prone position.

**Conclusions**

The coordinate transformations required for Monte Carlo simulation of radiation therapy using DOSZYZnrc and VMC++ codes have been generalized to account for the patient's orientation at CT scanning. In addition, corrections to the phantom generated by the ctcreate code were implemented to account for the DICOM *Image Orientation* tag. The transformations were validated by comparison with the Eclipse[TM] treatment planning system for all possible patient orientations.